\documentclass[paper=a4,abstracton,numbers=noenddot]{scrartcl}

\usepackage[english]{babel}

\usepackage{hyperref}
\hypersetup{%
    urlbordercolor = {1 1 1},
}
\usepackage{xspace}
\usepackage[intlimits]{amsmath}
\usepackage{amsthm}
\usepackage{amssymb}
\usepackage{units}
\usepackage{amsfonts,mathrsfs,bbm}
\usepackage{amsfonts,amssymb, bm}
\usepackage{mathtools}
\allowdisplaybreaks[1]
\usepackage[numbers]{natbib}
\usepackage{graphicx}
\usepackage{tikz}
\usepackage[T1]{fontenc}
\usepackage{lmodern}  

\usepackage[scaled=.92]{helvet}
\usepackage{courier}
\usepackage{subfig}
\usepackage{tikz-3dplot}
\usetikzlibrary{shapes,arrows}
\usepackage{pgfplots}
\pgfplotsset{every axis/.append style={thick}}
\pgfplotsset{every axis legend/.append style={cells={anchor=west},anchor=west}}
\pgfplotscreateplotcyclelist{mylines}{%
  {red,mark=*},
  {blue,mark=square},
  {green,mark=+},
  {black,mark=o},
  {violet,mark=otimes},
  {brown,mark=triangle},
  {cyan,mark=diamond},
  {orange,mark=10-pointed star},
  {magenta,mark=|}  }
\usetikzlibrary{external}

\theoremstyle{remark}
\newtheorem{remark}{Remark}

\usepackage[capitalise]{cleveref}
\crefformat{figure}{Fig.~#2#1#3} 
\Crefformat{figure}{Fig.~#2#1#3} 
\crefmultiformat{figure}{Figures~#2#1#3}{ and~#2#1#3}{, #2#1#3}{ and~#2#1#3} 
\crefformat{equation}{equation~(#2#1#3)} 
\Crefformat{equation}{Equation~(#2#1#3)} 
\labelcrefformat{equation}{(#2#1#3)}  
\crefformat{section}{Sect.~#2#1#3} 
\Crefformat{section}{Sect.~#2#1#3} 
\crefformat{listing}{Algorithm~(#2#1#3)} 
\Crefformat{listing}{Algorithm~(#2#1#3)} 

\newcommand{\mychangeEq}[1]{\color{black}#1\color{black}}
\newcommand{\mychange}[1]{\color{black}#1\color{black}\xspace}

\newcommand{\tikzfig}[5]{
    \centering
    \includegraphics{#5}
    \caption{#3}
    \label{#4}
}

\newcommand{\tikzsubfig}[5]{
    \centering
    \subfloat[#3]{\label{#4}\includegraphics{#5}}
}
\newcommand{\tikzsubfigNoCaption}[3]{
    \centering
    \includegraphics{#3}
}

\newcommand{\BEZ}{B{\'e}zier\xspace}

\newcommand{\AssemblyElemLoop}{Hybrid Gauss\xspace}

\newcommand{\AssemblyAddCut}{Disc.\,weighted quadrature\xspace}
\newcommand{\AssemblyNaive}{Weighted quadrature\xspace}

\newcommand{\R}{\mathbb{R}}

\newcommand{\myMat}[1]{\mathbf{#1}}

\newcommand\fromto[2]{\{ #1, \dots, #2 \}}
\DeclareMathOperator{\supp}{supp}
\newcommand{\Supp}[1]{\supp\{ #1\}}

\newcommand{\OfOrder}{\mathcal{O}}

\newcommand\pt[1]{\boldsymbol{#1}}

\newcommand\indexA{i} 		
\newcommand\indexB{j} 		
\newcommand\indexC{k} 		

\newcommand{\Bspline}{B} 		
\newcommand{\KV}{\varXi} 		
\newcommand{\KVRefined}{\tilde{\varXi}} 	
\newcommand\uu{\xi} 			
\newcommand\uuRefined{\tilde{\uu}} 	
\newcommand\vv{\uu} 			
\newcommand\UVsurf{\pt{\uu}} 	
\newcommand\uusurf{\uu_1} 		
\newcommand\vvsurf{\vv_2} 		
\newcommand\multi{m} 			

\newcommand\pu{p}			
\newcommand\pusurf{p_1}
\newcommand\pvsurf{p_2}

\newcommand{\CPara}{c}			

\newcommand\ndofs{n}
\newcommand\pdim{d}			

\newcommand{\trim}{t}

\newcommand{\domain}{\Omega}
\newcommand{\patchdomain}{\domain^{\textnormal{v}}}
\newcommand\indexSpan{s}	
\newcommand{\supportdomain}{\mathcal{S}^{\textnormal{v}}}
\newcommand{\supportdomainReg}{\mathcal{S}^{\textnormal{r}}}
\newcommand{\supportdomainCut}{\mathcal{S}^{\textnormal{t}}}

\newcommand{\visibledomain}{\patchdomain} 
\newcommand{\TrimCurve}{\pt{C}^{t}}     

\newcommand{\trialFct}{\Bspline}
\newcommand{\testFct}{\Bspline}
\newcommand{\quadWeight}{w}
\newcommand{\quadPoint}{x}
\newcommand{\quadRule}{\mathbb{Q}}
\newcommand{\quadIndexSet}{\mathcal{Q}}
\newcommand{\discCoeff}{{\uu}^{\textnormal{disc}}}

\newcommand\subDMatrix{\myMat{S}}

\author{    Benjamin Marussig \\ [1em]
  Institute of Applied Mechanics, \\Graz Center of Computational Engineering (GCCE),\\ Graz Univeristy of Technology,\\  Technikerstra\ss e 4/II 8010 Graz, Austria \\ \href{www.mech.tugraz.at}{\textcolor{blue}{www.mech.tugraz.at}},  \href{www.gcce.tugraz.at}{\textcolor{blue}{www.gcce.tugraz.at}}}

\title{Fast formation and assembly of isogeometric Galerkin matrices for trimmed patches}
\begin{document}

\maketitle
\thispagestyle{empty}
\begin{abstract}
This work explores the application of the fast assembly and formation strategy from \cite{Calabro2017a,Hiemstra2019a} to trimmed bi-variate parameter spaces.
Two concepts for the treatment of basis functions cut by the trimming curve are investigated:
one employs a hybrid Gauss-point-based approach, and the other computes discontinuous weighted quadrature rules.
The concepts' accuracy and efficiency are examined for the formation of mass matrices and their application to $L^2$-projection.
Significant speed-ups compared to standard element by element finite element formation are observed.
There is no clear preference between the concepts proposed.
While the discontinuous weighted scheme scales favorably with the degree of the basis, it also requires
additional effort for computing the quadrature weights.
The hybrid Gauss approach does not have this overhead, which is determined by the complexity of the trimming curve.
Hence, it is well-suited for moderate degrees, whereas discontinuous-weighted-quadrature has potential for high degrees, in particular, if the related weights are computed in parallel.
\end{abstract}



\section{Introduction}
\label{sec:Introduction}

Isogeometric analysis (IGA) has been introduced to overcome the profound inefficiencies in the conventional interaction of CAD and simulation tools \cite{Cottrell2009b,Hughes2005a}. 
The ground-breaking idea of IGA is to perform numerical simulations using CAD technology such as (non-uniform rational) B-splines. 
During the last decade, it has been demonstrated that this paradigm provides several computational benefits.
Indeed, IGA outperforms traditional simulations in most academic benchmarks \cite{Cottrell2007a,Cottrell2006a,Kiendl2009a_BM,Lipton2010a} and is nowadays generally recognized as a powerful alternative to the conventional analysis methodology.

The straightforward utilization of high-degree and high-continuity basis functions is an outstanding benefit of IGA.
Together with the concept of $k$-refinement, the resulting analysis features high robustness and accuracy w.r.t.~the degrees of freedom employed, see, e.g.~\cite{Beer2020,daVeiga2011a,evans2009a}.
This ability to perform high-order accurate simulations is, however, somewhat limited due to the computational cost, because the formation and assembly of the system matrix gets more involved with increasing polynomial degree $\pu$.
The state-of-the-art at the core of standard finite element codes is an element-wise assembly. 
IGA adopted this concept, and the introduction of {\BEZ extraction} \cite{Borden2010a,deBorst2017a} has provided a means to map between a smooth spline basis and an element-based representation.
Being compatible with conventional analysis routines has played an essential role for the acceptance and dispersion of IGA in the numerical analysis community and allowed a simple integration of this new paradigm into existing simulation software.
Yet, the corresponding cost for setting up the system matrix for a $C^{\pu-1}$-continuous $\pdim$-dimensional tensor product B-spline basis is \mychange{$\OfOrder\left(\pu^{3\pdim}\right)$} per degree of freedom \cite{Calabro2017a}.
Consequently, high-order analysis is doomed to be computationally expensive when a conventional matrix formation strategy is employed.

Reduced quadrature rules \cite{Auricchio2012aa,Hiemstra2017a,Hughes2010a,Johannessen2017a,Schillinger2014a} improve the efficiency of the numerical integration on the element-level.
However, the element-wise assembly by itself limits the computation cost to $\OfOrder\left(\pu^{2\pdim}\right)$ \cite{Calabro2017a}.
\emph{Sum factorization} \cite{Antolin2015a,Orszag1980a} approaches this threshold by rearranging the computations to exploit the tensor product structure of the B-spline basis. \mychange{The resulting cost using element-wise Gaus quadrature is $\OfOrder\left(\pu^{2\pdim+1}\right)$.} 
As shown in \cite{Bressan2019a}, sum factorization with Gauss quadrature can even yield a computational complexity of  $\OfOrder\left(\pu^{\pdim+2}\right)$, 
when used globally and not on an element level. 
\emph{Weighted quadrature} introduced in \cite{Calabro2017a} is a new integration technique that also drops the element perspective.  
A quadrature rule is set up for each test function by incorporating the test function into the quadrature weights.
The outstanding feature of the resulting quadrature is its independence on the degree $\pu$ for spline with maximal smoothness, i.e., $C^{\pu-1}$.
The application of weighted quadrature to non-uniform spline with mixed continuity is addressed in \cite{Hiemstra2019a}.
Furthermore, the integration of sum factorization and weighted quadrature into a \emph{row-based assembly} strategy results in the fast formation and assembly approach detailed in \cite{Calabro2017a,Hiemstra2019a}. 
These three ingredients -- row assembly, sum factorization, and weighted quadrature -- reduce the computational cost to $\OfOrder\left(\pu^{\pdim+1}\right)$.

The treatment of trimmed patches has been denoted as an open challenge for this fast formation and assembly approach at the recent INdAM Workshop on Geometric Challenges in Isogeometric Analysis.
Trimming is an essential concept for representing complex geometries with tensor product B-splines. 
The main component involved is the \emph{trimming curve}, which is specified in the parameter space and restricts the visible part
of the spline object to a subregion. 
The resulting trimmed space consists of interior, exterior, and cut basis functions. 
The latter introduces several computational challenges as detailed in \cite{marussig2017a}.
For example, the realization of a high-order accurate numerical integration scheme is far from trivial even in the classical finite element setting, see e.g.~\cite{Fries2015a}.
Moreover, cut basis functions do not follow the tensor product structure of the basis anymore.
In order words, trimming violates an essential property for the fast formation process. 
Hence, the question arises if
\mychange{a significant reduction of the computational costs is restricted to the non-trimmed case.}

This paper provides extensions to the fast formation and assembly approach presented in \cite{Calabro2017a,Hiemstra2019a} that allow the analysis of trimmed domains. 
In particular, the integration over cut basis functions is investigated using either a (i) Gauss quadrature or a (ii) weighted quadrature approach. 
Their performance and accuracy are compared for the mass matrix formation of trimmed bi-variate spaces with different complexity.
Both concepts maintain the optimal approximation order and significantly improve efficiency compared to a standard element based Gaussian assembly.
From a conceptional point of view, the extension to tri-variate spaces is straightforward, but the implementation gets more involved due to the increased topological complexity of the cut elements.

The paper is \mychange{structured} as follows: 
\Cref{sec:fastformation} outlines the fast formation and trimming concepts and highlights the contradiction of their underlying ideas. 
The proposed approaches to overcome this barrier are presented in \cref{sec:integration} and they are compared by numerical experiments in \cref{sec:NumericalResults}.
The paper closes with concluding remarks.
\section{The discrepancy between weighted quadrature and trimming}

This section provides short introductions to weighted quadrature, sum factorization, and trimmed patches, which are
the preliminaries for the proposed approach detailed subsequently. 
First, the main aspects presented in \cite{Calabro2017a,Hiemstra2019a} are recapitulated.
Then, the fundamentals of trimmed spaces and the challenge of deriving a fast formation technique for them is outlined.

\subsection{Fast formation by weighted quadrature and sum factorization}
\label{sec:fastformation}

\mychange{Let us recapitulate some essential properties of the basis functions used.}
A \emph{B-spline} $\Bspline_{\indexA,\pu}$ is described by piecewise polynomial segments of degree $\pu$.
The continuity between them is specified by the {knot vector}~$\KV$, which is a non-decreasing sequence of parametric coordinates~{$\uu_\indexB \leqslant \uu_{\indexB+1}$}.  
The values of these {knots} $\uu_\indexB$ define the location where adjacent segments join. 
The continuity at these {breakpoints} is $C^{\pu-\multi}$, with $\multi$ denoting the multiplicity of the corresponding knot value, i.e., $\uu_{\indexB} = \uu_{\indexB+1} = \dots = \uu_{\indexB+\multi-1}$. 
The {knot span} $\indexSpan$ refers to the half-open interval $\left[\uu_{\indexSpan}, \uu_{\indexSpan+1}\right)$, and if its size is non-zero, it marks an {element}. 
Furthermore, the knot vector $\KV$ defines an entire set of linearly independent B-splines~$\{\Bspline_{\indexA,\pu}\}_{\indexA=0}^{\ndofs}$ on the \mychange{parametric domain} $\domain$.
Each $\Bspline_{\indexA,\pu}$ has local support, $\supp{ \{\Bspline_{\indexA,\pu} \} }$, specified by the knots $\fromto{\uu_{\indexA}}{\uu_{\indexA+\pu+1}}$, and each knot span $\indexSpan$ contains $\pu+1$ non-zero B-splines. 
Bi-variate basis functions are obtained by computing the tensor product of uni-variate B-splines $\Bspline_{\indexA_1,\pusurf}$ and $\Bspline_{\indexA_2,\pvsurf}$ of degrees $\pusurf$ and $\pvsurf$, which are defined by separate knot vectors $\KV_1$ and $\KV_2$  for the parametric directions $\uusurf$ and $\vvsurf$, respectively.
This can be generally expressed as
\begin{align}
    \label{eq:tensorProductBspline}
    \Bspline_{\boldsymbol{\indexA}}(\UVsurf) = \Bspline_{\indexA_1,\dots,\indexA_\pdim}(\uu_1,\dots,\uu_\pdim)=\prod_{\indexC=1}^\pdim \Bspline_{\indexA_\indexC}(\uu_\indexC).
\end{align}

In this paper, the focus lies on \emph{weighted quadrature} for the formation of mass matrices. 
Hence, the aim is to derive an efficient and accurate evaluation of the following uni-variate integral
\begin{align}
    \label{eq:integralOfInterest}
    \int_\domain \testFct_\indexA(\uu) \trialFct_\indexB(\uu) \CPara(\uu) d\uu.
\end{align}
$\testFct_\indexA(\uu)$ and $\trialFct_\indexB(\uu)$ are test and trial functions and $\CPara(\uu)$ is determined by the geometry mapping.
In general, numerical quadrature rules $\quadRule$ are designed to be exact for the case that $\CPara(\uu)=1$, and they provide quadrature weights $\quadWeight_\indexC$, which allow the expression of \cref{eq:integralOfInterest} as a sum over corresponding quadrature points $\quadPoint_\indexC$, i.e.,
\begin{align}
    \label{eq:quadRule}
    \quadRule = \sum_\indexC \testFct_\indexA(\quadPoint_\indexC) \trialFct_\indexB(\quadPoint_\indexC)  \quadWeight_\indexC \coloneqq \int_\domain \testFct_\indexA(\uu) \trialFct_\indexB(\uu) d\uu.
\end{align}
The novelty of {weighted quadrature} is that a quadrature rule $\quadRule_\indexA$ is designed for each test function $\testFct_\indexA$ by incorporating the test function into the quadrature weights
\begin{align}
    \label{eq:weightedQuadRule}
    \quadRule_\indexA = \sum_\indexC \trialFct_\indexB(\quadPoint_\indexC)  \quadWeight_{\indexC,\indexA} \coloneqq \int_\domain \trialFct_\indexB(\uu)  \left(\testFct_\indexA(\uu)  d\uu \right).
\end{align}
The computation of the weights $\quadWeight_{\indexC,\indexA}$ requires an adequate distribution of the quadrature points $\quadPoint_\indexC$. 
Here, the procedure proposed in \cite{Hiemstra2019a} is employed, and the reader is referred to this publication for details. It determines the minimal number of quadrature points for each element of a non-uniform knot vectors with mixed continuity. 
Subsequently, a uniform distribution within the elements' interior is employed so that no $\quadPoint_\indexC$ coincides with a knot $\uu_\indexA$.

\begin{figure}[ht]
    \centering
    \subfloat[][Bi-degree $\pu=2$]{\includegraphics[width=0.45\textwidth]{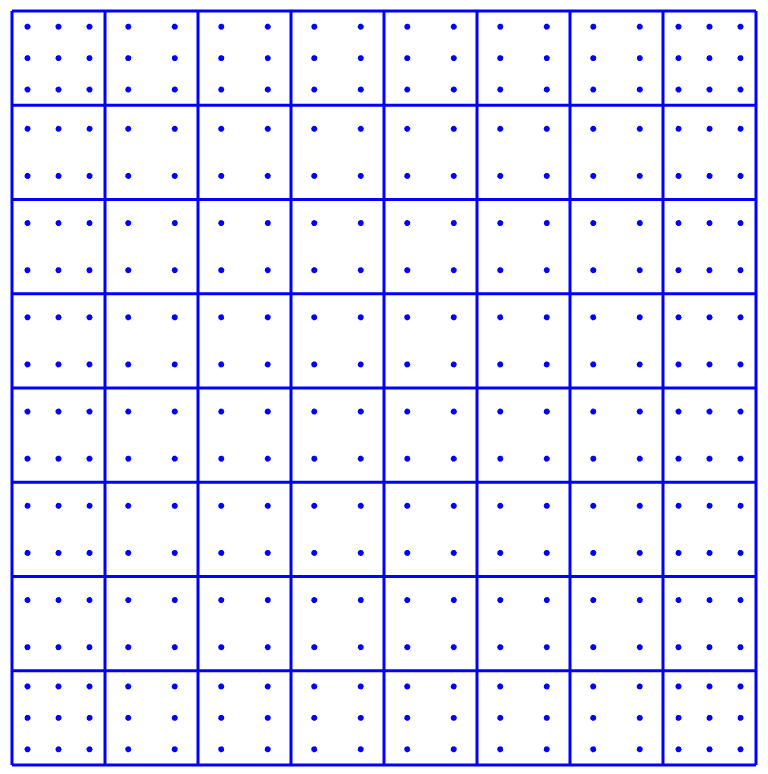}\label{fig:ParameterSpaceWQh8p2c}}
    \hfill
    \subfloat[][Bi-degree $\pu=6$]{\includegraphics[width=0.45\textwidth]{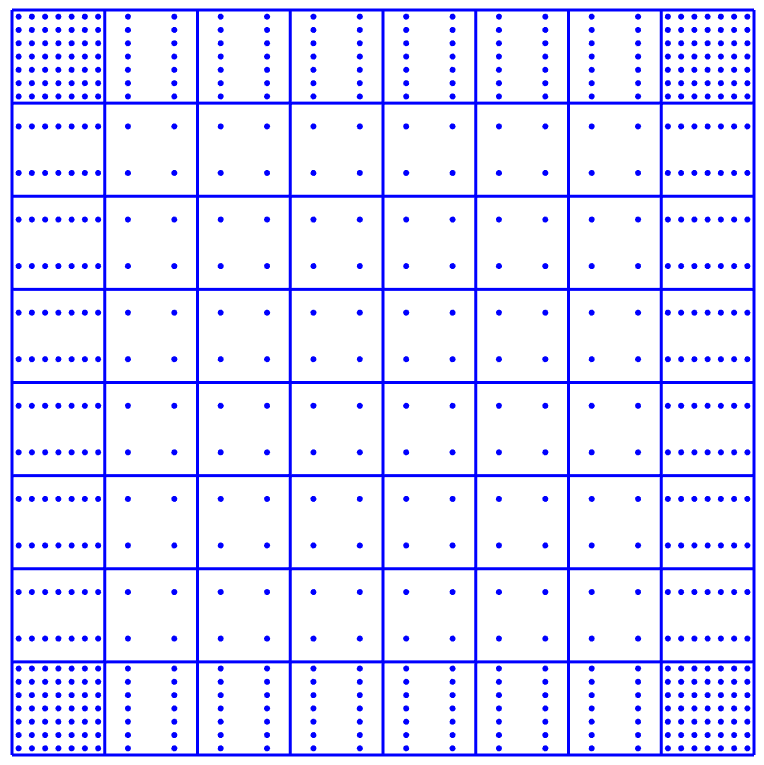}\label{fig:ParameterSpaceWQh8p6c}}
    \caption{Weighted quadrature points for two bi-variate bases with the same number of elements but with a different polynomial degree.}
    \label{fig:ParameterSpaceWQ}
\end{figure}

Once the position of the quadrature points $\quadPoint_\indexC$ is fixed, 
the corresponding weights can be computed by the following system of equations
\begin{align}
    \begin{aligned}
        \label{eq:weightedQuadRuleSystem}
        \quadRule_\indexA\left(\trialFct_{\indexB_1}\right) &=&& \sum_{\indexC\in\quadIndexSet_\indexA} \trialFct_{\indexB_1}(\quadPoint_\indexC)  \quadWeight_{\indexC,\indexA}&&\coloneqq \int_\domain \trialFct_{\indexB_1}(\uu)  \left(\testFct_\indexA(\uu)  d\uu \right) \\
        &\vdotswithin{=}&&  &&\vdotswithin{:=}\\
        \quadRule_\indexA\left(\trialFct_{\indexB_n}\right) &=&& \sum_{\indexC\in\quadIndexSet_\indexA}  \trialFct_{\indexB_n}(\quadPoint_\indexC)  \quadWeight_{\indexC,\indexA}&&\coloneqq \int_\domain \trialFct_{\indexB_n}(\uu)  \left(\testFct_\indexA(\uu)  d\uu \right) 
    \end{aligned}
\end{align}
where $\indexB_1,\dots,\indexB_n$ are the indices of all trial functions whose support overlaps with the one of the current test function, $\supp{ \{\Bspline_{\indexA} \} }$, and the index set $\quadIndexSet_\indexA$ refers to all quadrature points that lie within $\supp{ \{\Bspline_{\indexA} \} }$. 
Solving \labelcref{eq:weightedQuadRuleSystem} by QR-factorization yields $\quadWeight_{\indexC,\indexA}$ for $\forall k\in \quadIndexSet_\indexA$, and the remaining quadrature weights are set to zero. 
The solution of \labelcref{eq:weightedQuadRuleSystem} maybe not unique, and the system of equations can be weighted to improve positivity and boundedness of the weights as detailed in \cite{Hiemstra2019a}.  
In the case of multi-variate tensor product basis functions, quadrature rules $\quadRule_{\indexA_1},\dots,\quadRule_{\indexA_\pdim}$ are computed for each parametric direction.
\Cref{fig:ParameterSpaceWQ} shows the quadrature layout for two bases of degree 2 and 6, respectively. 
Note that the number of quadrature points for the inner elements does not change.

\emph{Sum factorization} takes advantage of the tensor product structure of the quadrature rule and the spline basis by a reordering of the numerical operations such that only uni-variate quadrature rules are required.
Consider the computation of an entry of a mass matrix for a bi-variate basis
\begin{align}
    \label{eq:massMatrixEntry}
    m_{\boldsymbol{\indexA},\boldsymbol{\indexB}} = \int_\domain \Bspline_{\boldsymbol{\indexA}} (\UVsurf) \Bspline_{\boldsymbol{\indexB}}(\UVsurf) \CPara(\UVsurf) d\UVsurf.  
\end{align}
Starting from \labelcref{eq:tensorProductBspline}, \cref{eq:massMatrixEntry} can be expressed as a recursion of uni-variate integrals
\begin{align}
    \label{eq:massMatrixEntryReordered}
    m_{\boldsymbol{\indexA},\boldsymbol{\indexB}} = 
    \int_{\domain_1} \Bspline_{\indexA_1} (\uu_1) \Bspline_{\indexB_1}(\uu_1) \times
    \left[
        \int_{\domain_2} \Bspline_{\indexA_2} (\uu_2) \Bspline_{\indexB_2}(\uu_2) 
        \CPara(\uu_1,\uu_2) 
        d\uu_2
        \right]
    d\uu_1.  
\end{align}
Consequently, it is straightforward to apply the weighted quadrature rules $\quadRule_{\indexA_1},\quadRule_{\indexA_2}$ associated with a bi-variate test function $\testFct_{\boldsymbol{\indexA}}$. 
Finally, it is noted that the equations have to be pulled back to the parametric domain \cite{Antolin2015a,Hiemstra2019a} to exploit the structure of \cref{eq:massMatrixEntryReordered}.

\subsection{Trimmed patches}
\label{sec:method}

Tensor product B-splines allow full control over the continuity and degree of the basis, but they possess an intrinsic four-sided topology limiting their ability to represent arbitrary domains.
Trimming provides a remedy to this restriction by defining the valid area~$\visibledomain$ independent from the basis' structure. 
In particular, curves in the parametric space specify $\visibledomain$, and the curve's orientation determines the interior and exterior domain. 
Usually, these \emph{trimming curves} $\TrimCurve$ are represented by B-splines as well \cite{marussig2017a}, but this choice is more or less \mychange{free}. 

The presence of $\TrimCurve$ divides the basis into three different functions types based on the overlap of the support with the valid domain, i.e., $\supportdomain_{\boldsymbol{\indexA}} \coloneqq  \Supp{ \Bspline_{{\boldsymbol{\indexA}}} }  \cap \mychangeEq{\overline{\patchdomain}}$.
That is, a B-spline $\Bspline_{{\boldsymbol{\indexA}}}$ is classified as:
\begin{itemize}
    \item \emph{Exterior} if $\supportdomain_{\boldsymbol{\indexA}} = \emptyset$, 
    \item \emph{Interior} if $\supportdomain_{\boldsymbol{\indexA}} = \Supp{ \Bspline_{{\boldsymbol{\indexA}}} }$, 
    \item \emph{Cut} if  \mychange{$0 < \left|\supportdomain_{\boldsymbol{\indexA}}\right| <\left| \Supp{ \Bspline_{{\boldsymbol{\indexA}}} }\right|$,} 
\end{itemize}
\mychange{where $\left|\cdot\right|$ denotes the Lebesgue measure in $\R^\pdim$.}
\Cref{fig:basisFunctionTypes} illustrated these different classes for a bi-variate cubic basis.
In the context of analysis, exterior basis functions can be neglected from the system of equations, and interior ones can be treated as usual.
Cut basis functions, however, induce profound numerical challenges regarding the application of boundary conditions,
the conditioning of system matrices, and accurate integration. 
\begin{figure}[b!]
    \tikzfig{tikz/TrimmedSpaceExample.tex}{0.5}{Trimmed cubic bi-variate basis with the trimming curve $\TrimCurve$ specifying the valid domain $\visibledomain$ (gray). The resulting B-splines types are interior (green), cut (red), or exterior (yellow) based on the overlap of the support, $\Supp{ \Bspline_{\indexA_1,\indexA_2} }$, with $\visibledomain$.}{fig:basisFunctionTypes}{FigTrimmedSpaceExample}
\end{figure}
Addressing these aspects is far from trivial, even in the conventional low-order finite element setting.
The loss of the tensor product structure and the requirement of higher-order accuracy complicates the situation further for the fast formation outlined in \cref{sec:fastformation}.

In the following, the focus is on integrating cut basis functions. The application of boundary conditions is not an issue in this paper because only the formation of mass matrices and $L^2$-projection is considered.
The extended B-spline concept is employed to address the conditioning aspect. 
This procedure is independent of the assembly and formation process, and hence, it is not described here. The interested reader is referred to \cite{Hoellig2003b,marussig2018a,marussig2016a}.

\section{Integration of cut basis functions}
\label{sec:integration}

\mychange{A trimming curve $\TrimCurve$ introduces an arbitrarily located jump discontinuity within the parameter space.}
Simply integrating over this interface or neglecting quadrature points that lie outside of the valid domain $\visibledomain$ will evidently lead to incorrect results. 
Consequently, the numerical integration scheme has to account for these arbitrarily located discontinuities.
In the case of weighted quadrature, this circumstance affects not only the correct representation of the integration domain $\supportdomain_{\boldsymbol{\indexA}}$ of a cut B-spline $\Bspline_{\boldsymbol{\indexA}}$ but also the computation of its quadrature rules $\quadRule_{\indexA_1}$ and $\quadRule_{\indexA_2}$.
Moreover, sum factorization cannot be applied because $\supportdomain_{\boldsymbol{\indexA}}$ does not follow a tensor product structure in general. 

However, we can split the domain $\supportdomain_{\boldsymbol{\indexA}}$ into a \emph{regular} part $\supportdomainReg_{\boldsymbol{\indexA}}$, which follows the tensor product structure (at least on the element-level), and a \emph{trimmed} part $\supportdomainCut_{\boldsymbol{\indexA}}$, which consists of all elements cut by the trimming curve.
The integral over a cut basis function can be written as
\begin{align}
    \label{eq:domainSplitting}
    \int_{\supportdomain_{\boldsymbol{\indexA}}}    \Bspline_{\boldsymbol{\indexA}}(\UVsurf) d\UVsurf = 
    \int_{\supportdomainReg_{\boldsymbol{\indexA}}} \Bspline_{\boldsymbol{\indexA}}(\UVsurf) d\UVsurf + 
    \int_{\supportdomainCut_{\boldsymbol{\indexA}}} \Bspline_{\boldsymbol{\indexA}}(\UVsurf) d\UVsurf. 
\end{align}
In the following, the numerical integration of ${\supportdomainCut_{\boldsymbol{\indexA}}}$ employs a standard element-wise assembly procedure and \cref{sec:TreatmentOfCutElements} lists various concepts for the distribution of quadrature points.  
Furthermore, this work investigates two options for the treatment of the remaining regular part $\supportdomainReg_{\boldsymbol{\indexA}}$: In \cref{sec:HybridQuadrature}, it is integrated using Gaussian quadrature, while \cref{sec:DiscontinuousWeightedQaudrature} derives a weighted quadrature rule that acknowledges the presence of the trimming curve within the support.

\subsection{Treatment of cut elements}
\label{sec:TreatmentOfCutElements}

The fast formation and assembly approach opens the path towards efficient higher-order simulation. 
Hence, it is of utmost importance that the analysis, and therefore the integration of cut elements, is performed with higher-order accuracy.
%
Luckily, there is a substantial body of literature on this topic because integrating over elements cut by an arbitrary interface is a canonical problem in various analysis schemes such as fictitious domain methods, extended finite element approaches, and the simulation with trimmed spline geometries.
The proposed concepts can be broadly divided into strategies that
(i) set  up tailored integration rules or
(ii) decompose cut elements into sub-elements, which then employ standard quadrature rules. 
For the latter type, several works \cite{Antolin2019a,cheng2010a,Fries2015a,Kudela2013Masterthesis,Kudela2016a,Kudela2015a,legay2005a} have demonstrated that the introduction of curved sub-elements with high degree, whose edges or faces capture the interface, allows for higher-order accurate integration with a modest number of integration points.
However, generating these sub-elements is a challenge on its own, and its complexity is determined by the dimension of the domain and the shape of the interface. 
An alternative is the construction of tailored integration rules for a cut element, see e.g.~\cite{mousavi2011aa,muller2013a,Nagy2014a}.
These strategies usually obtain integration weights by solving a system of moment-fitting equations.
Utilizing Lasserre's theorems \cite{lasserre1998a}, integrals over the boundaries of the cut elements provide the right-hand side required.
M\"{u}ller et al.~\cite{muller2013a} showed that this concept achieves higher-order accurate integration over curved three-dimensional domains.

\begin{figure}[ht]
    \centering
    \subfloat[][Bi-degree $\pu=2$]{\includegraphics[width=0.45\textwidth]{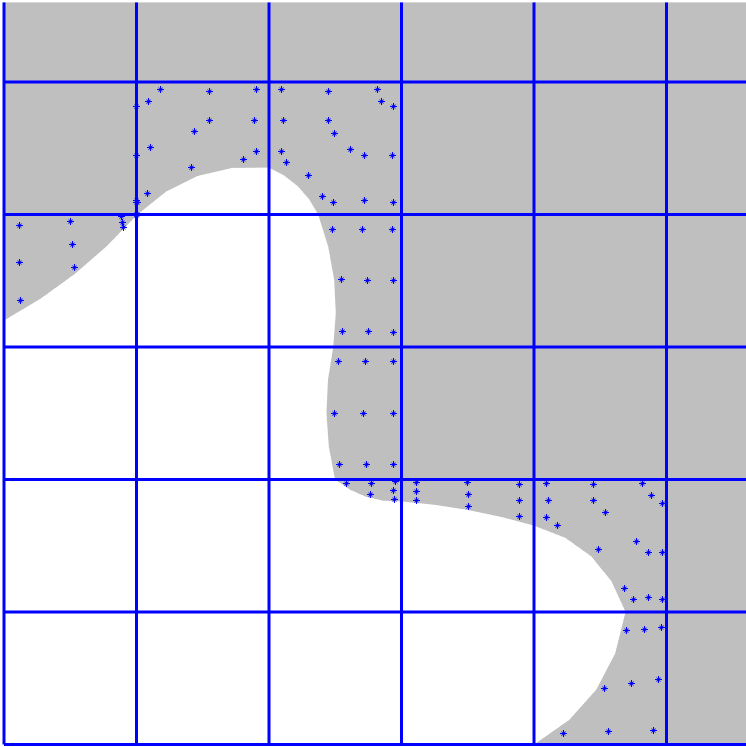}\label{fig:ParamterSpaceTrimmedGQh8p2CutElCloseUpc}}
    \hfill
    \subfloat[][Bi-degree $\pu=6$]{\includegraphics[width=0.45\textwidth]{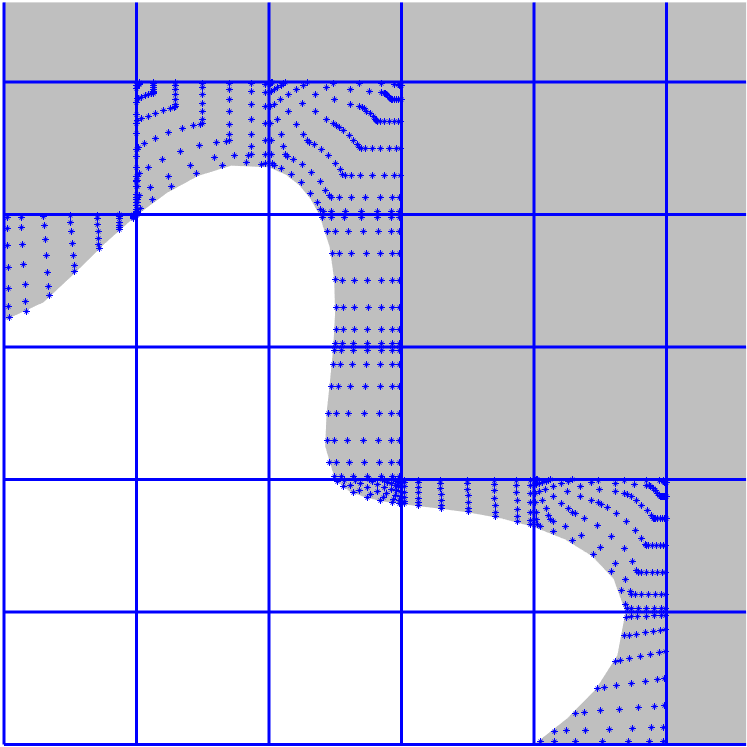}\label{fig:ParamterSpaceTrimmedGQh8p6CutElCloseUpc}}
    \caption{Higher-order accurate distribution of Gaussian quadrature points within cut elements for different degree.}
    \label{fig:ParameterSpaceTrimmedGQ}
\end{figure}

Herein, the decomposition-based approach detailed in \cite{Fries2015a,Schoellhammer2020a} is utilized.
It represents the valid domain $\visibledomain$ in the parameter space by a set of higher-order Lagrange elements, which are aligned with the parameter space as well as the trimming curve $\TrimCurve$. Furthermore, they have the same polynomial degree as the underlying B-spline basis so that the representation quality of $\TrimCurve$ matches the approximation power sought for the analysis. 
The mappings of these Lagrange elements distribute Gauss quadrature points within cut elements, as shown in \cref{fig:ParameterSpaceTrimmedGQ}.
It is worth noting that this particular choice for the integration of cut elements is not essential for the ideas of this work.
Any integration scheme could be employed, as long as it allows for higher-order accurate integration of trimmed domains. 
Therefore, a detailed discussion on this part of the overall integration of cut B-splines is omitted.

\subsection{Hybrid Gauss approach}
\label{sec:HybridQuadrature}

\begin{figure}[h]
    \centering
    \subfloat[][Degree 2]{\includegraphics[width=0.45\textwidth]{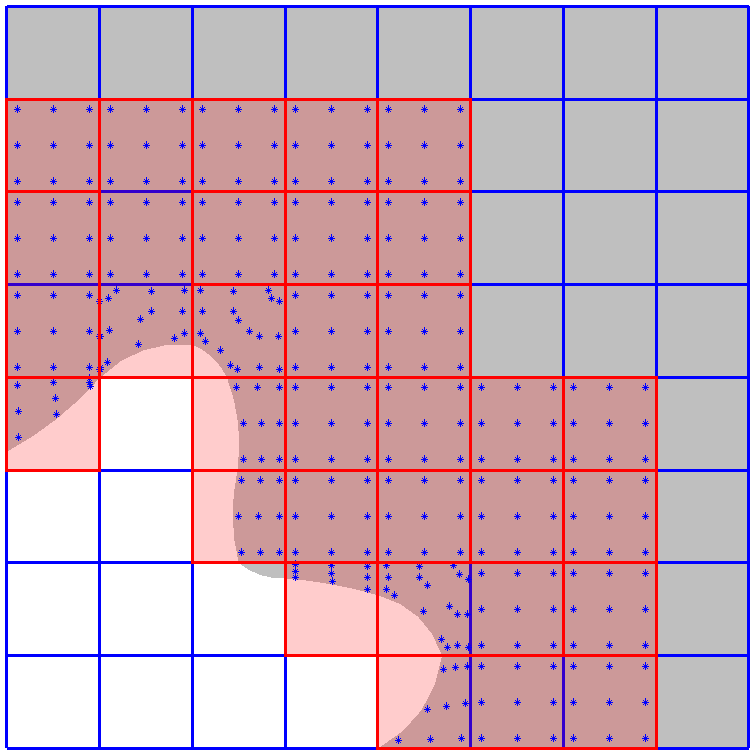}\label{fig:ParamterSpaceTrimmedGQh8p2SupportCutTestOnlyc}}
    \hfill
    \subfloat[][Degree 6]{\includegraphics[width=0.45\textwidth]{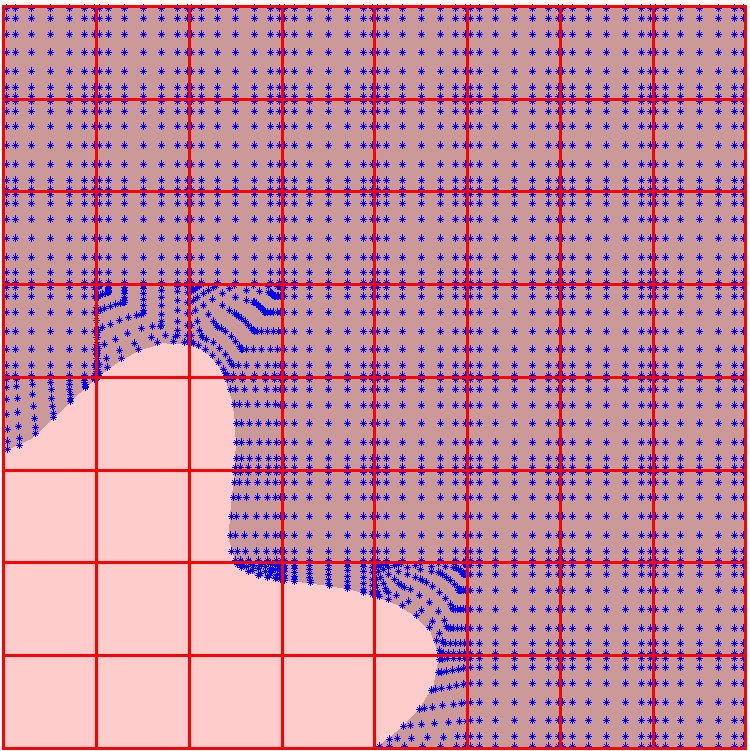}\label{fig:ParamterSpaceTrimmedGQh8p6SupportCutTestOnlyc}}
    \caption{Gaussian quadrature for the entire valid support $\supportdomain$ of all cut basis functions. The support of cut basis functions is highlighted in red.}
    \label{fig:ParameterSpaceTrimmedGQSupportCutTest}
\end{figure}

The perhaps most straightforward way to integrate over the regular portion  $\supportdomainReg_{\boldsymbol{\indexA}}$ of a cut basis function is to employ standard Gauss quadrature using a conventional element-wise assembly. In contrast to the cut elements of  $\supportdomainCut_{\boldsymbol{\indexA}}$, the evaluations during the formation can exploit the tensor product structure. Hence, it is beneficial to implement separate routines for  $\supportdomainReg_{\boldsymbol{\indexA}}$ and  $\supportdomainCut_{\boldsymbol{\indexA}}$.
As a result, a cut B-spline is treated by a hybrid Gauss-quadrature-based formation concept. 
Regarding the entire trimmed space, Gauss points are restricted to the vicinity of the trimming curve. 
Yet, they propagate further into the interior when the degree increases \mychange{since the supports of the trimmed B-splines increase.}  
Consider the example shown in \cref{fig:ParameterSpaceTrimmedGQSupportCutTest}, all elements are affected in the degree 6 case.
Nevertheless, the portion of the computational cost for setting up the entire system of equations decreases with the fineness of the parameter space.

\begin{remark}
    Since the elements of $\supportdomainReg_{\boldsymbol{\indexA}}$ follow the tensor product structure, sum factorization with \mychange{Gauss points $\pt{\quadPoint}_\indexC$ may be used by setting the precomputed coefficient $\CPara(\pt{\quadPoint}_\indexC)$ of \cref{eq:massMatrixEntry} to zero if $\pt{\quadPoint}_\indexC\notin\supportdomainReg_{\boldsymbol{\indexA}}$.}
    This option is, however, not considered here. 
\end{remark}

\subsection{Discontinuous weighted quadrature}
\label{sec:DiscontinuousWeightedQaudrature}

The domain splitting \labelcref{eq:domainSplitting} allows us to extract those elements that follow the tensor product structure. Hence, we can apply sum factorization there, as noted in the previous section. In the context of weighted quadrature, however, we cannot set the coefficients \mychange{$\CPara(\pt{\quadPoint}_\indexC)$ in \cref{eq:massMatrixEntryReordered} to zero if the quadrature point $\pt{\quadPoint}_\indexC\notin\supportdomainReg_{\boldsymbol{\indexA}}$.}
In contrast to Gauss quadrature, the weighted rules take advantage of the continuity between elements. Thus, the zero coefficients introduce artificial discontinuities {within} the domain of the quadrature, which leads to an error in the evaluation of the integral.   

The construction of \emph{discontinuous weighted quadrature} is proposed to resolve this problem.
The idea is to incorporate the information on the location $\discCoeff$ of these artificial discontinuities into the quadrature weights. 
As a result, the numerical integration on one side of $\discCoeff$ can be treated independently from the other.
Each $\discCoeff$ marks a knot adjacent to a cut element. 
Thus, the trimming curve determines the number of $\discCoeff$ required.
The tools for deriving related quadrature rules are the fact that quadrature is a linear operator and the properties of knot insertion.

Knot insertion denotes the refinement of a B-spline object by adding knots $\uuRefined$ into its knot vector $\KV$.
This procedure leads to nested spline spaces, and 
the {subdivision matrix} $\subDMatrix \: : \; \mathbb{R}^{\ndofs+1} \mapsto \mathbb{R}^{\tilde{\ndofs}+1}$ encodes the coefficients of the initial coarse representation and the refined one.
Considering quadrature weights, we obtain the relation
\begin{align}
	\label{eq:knotInsertion1}
	\tilde{\quadWeight}_i = \sum_{j=0}^{\ndofs} \subDMatrix_{ij}  \; \quadWeight_j && \text{for } i=0,\ldots , \tilde{\ndofs}
\end{align}
where $\tilde{w}_i$ refers to quadrature weights computed for the refined basis functions.
If only one knot is inserted, i.e., $\KVRefined = \KV \cup \uuRefined$, where $\uuRefined \in [\uu_\indexSpan,\uu_{\indexSpan+1})$, the non-zero entries of $ \subDMatrix$ are determined by
\begin{align}
	&
	\label{eq:knotInsertion}
	\begin{cases}
		 \mathbf{S}(\indexC,\indexC-1) &=  1-\alpha_\indexC 	\\
		\mathbf{S}(\indexC,\indexC) &=  \alpha_\indexC
	\end{cases}
	&
	&
   \alpha_\indexC =	\left\{ \begin{array}{c l}
			1  &  \indexC \leqslant \indexSpan-\pu \\
			\frac{\uuRefined - \uu_\indexC}{\uu_{\indexC+\pu} - \uu_\indexC } & \indexSpan-\pu+1  \leqslant \indexC \leqslant \indexSpan\\
			0  &  \indexC \geqslant \indexSpan+1 \\
			\end{array} 
		\right.
		&
\end{align}
Multiple knots can be inserted by repeating this process, and the multiplication of the individual single-knot matrices yields the overall subdivision matrix.

Using \mychange{the} relations provided by $\mathbf{S}$, the construction of discontinuities weighted quadrature rules for each artificial discontinuities $\discCoeff$ goes as follows:
\begin{enumerate}
    \item Knot insertion at $\discCoeff$ so that the uni-variate basis become $C^{-1}$ continuous there, and storing of the corresponding $\mathbf{S}$.
    \item Determination of the minimal number of quadrature points $n_{min}$ for the refined basis \cite{Hiemstra2019a}, and addition of new nested quadrature points if necessary.
    \item Computation of $\tilde{\quadWeight}$ for basis functions of the refined basis using \cref{eq:weightedQuadRuleSystem}.
    \item Multiplication of $\tilde{\quadWeight}$ by $\mathbf{S}^{\textnormal{T}}$ to obtain $\quadWeight$ for the initial basis.
\end{enumerate}
\Cref{fig:DWQConstruction} illustrates these steps for a single cut basis function.
Using nested quadrature points in step 2 of the construction allows the reuse of all coefficients set up for the weighted quadrature for interior B-splines during the integration process.
The resulting weights $\quadWeight$ of the initial and nested quadrature points are obtained by a linear combination of weights $\tilde{\quadWeight}$ that account for the discontinuity at $\discCoeff$. 
Thus, we can set all coefficients on one side of $\discCoeff$ to zero without affecting the integral on the other side.
\begin{figure}[t]
    \centering
    \tikzfig{tikz/DWQConstructionOneFct.tex}{1.45}{Discontinuous weighted quadrature (DWQ) for the cubic B-spline $\Bspline_{3,3}$ cut at the trimming position $\trim$. (a) Conventional weighted quadrature (WQ) points (black dots) for $\Bspline_{3,3}$. (b) Quadrature layout with additional quadrature points (white) for WQ of the refined discontinuous $\tilde{\Bspline}_{\indexB,3}$ associated to $\Bspline_{3,3}$. (c) Linear combination of the refined discontinuous WQ rules to obtain the DWQ for $\Bspline_{3,3}$. In (a,c), the points' height indicates the related weight value.}{fig:DWQConstruction}{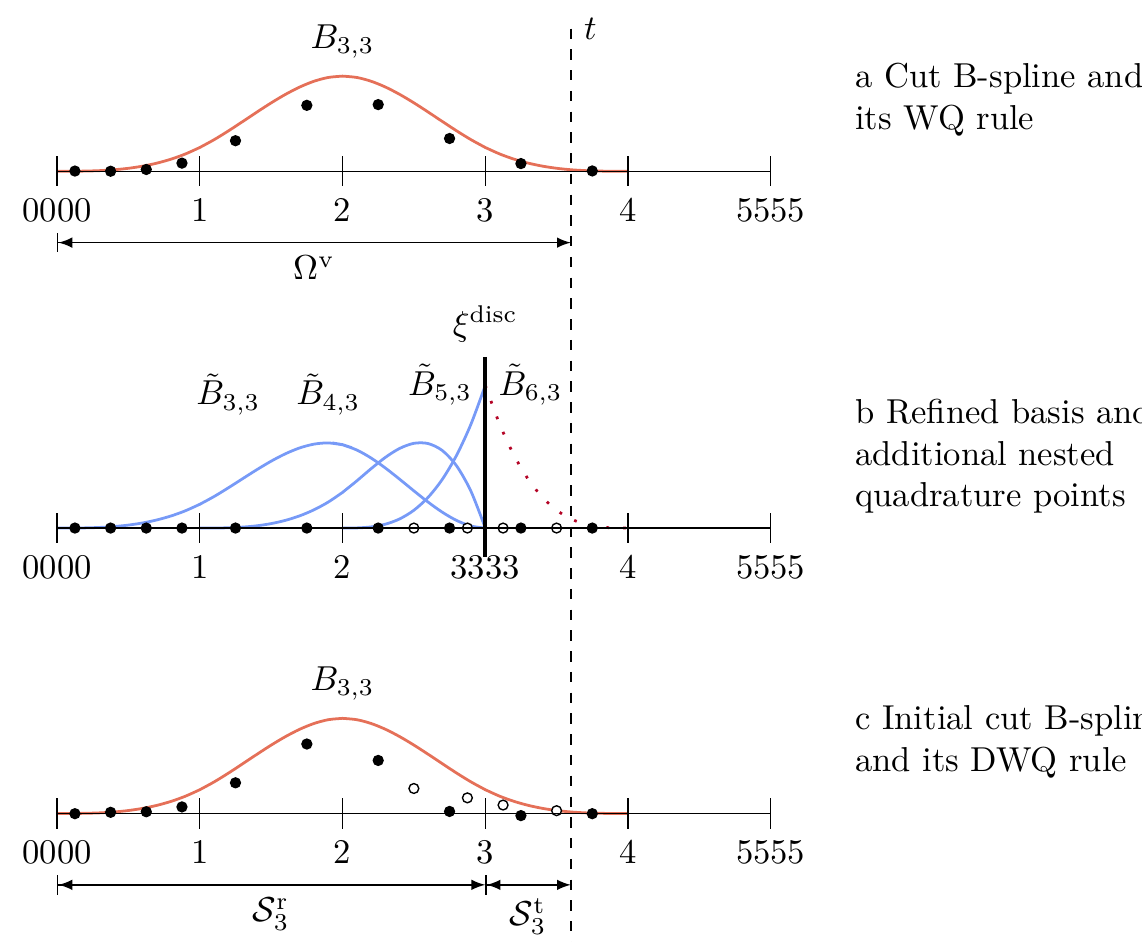}
\end{figure}

\mychange{%
\begin{remark}
    The nested quadrature points are uniformly distributed between the initial ones. The number of added points per interval is chosen such that the distance between between them gets as large as possible. 
\end{remark}
}

In the context of sum factorization, only discontinuous weighted quadrature points within $\supportdomainReg_{\boldsymbol{\indexA}}$ are of interest.
In a general trimming situation, these points do not cover all elements of $\supportdomainReg_{\boldsymbol{\indexA}}$.
The remaining parts are integrated by Gaussian quadrature.
Consider the example shown in \Cref{fig:TrimmedSpaceDiscWQExampleB}. 
Note that the artificial discontinuity restricts the Gauss points to the vicinity of the trimming curve, independent of the degree.
\begin{figure}[ht]
    \tikzfig{tikz/TrimmedSpaceDiscWQExampleB.tex}{0.36}{All quadrature points of a cut B-spline of bi-degree 6: above the artificial discontinuity $\discCoeff_2$ discontinuous weighted quadrature (DWQ) is employed, while Gauss points are used below. The white dots mark the nested quadrature points added during the construction.}{fig:TrimmedSpaceDiscWQExampleB}{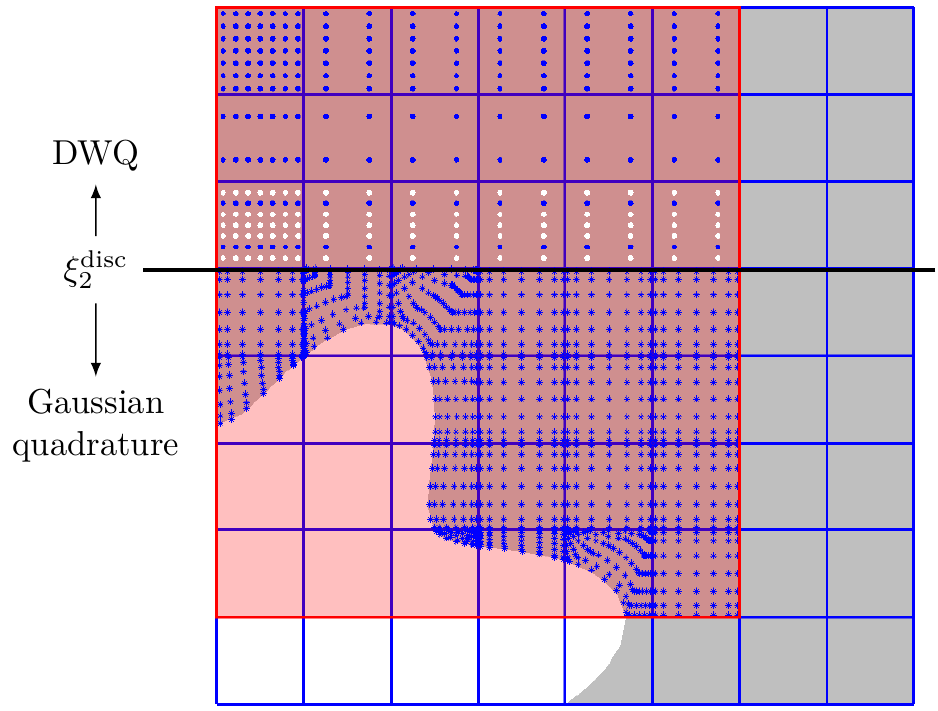}
\end{figure}

\begin{remark}
    Every artificial discontinuity $\discCoeff$ increases the number of required quadrature points in the elements adjacent to $\discCoeff$. Hence, the advantage of providing a discontinuous weighted quadrature rule for a cut basis functions decreases with the number of $\discCoeff$. In the present implementation, an individual quadrature is constructed only if not more than one $\discCoeff$ for each parametric direction is present in the basis function's support.
\end{remark}

\section{Numerical results}
\label{sec:NumericalResults}

\begin{figure}[t]
    \centering
    \subfloat[fig:DomainLine][Line]{\includegraphics[width=0.3\textwidth]{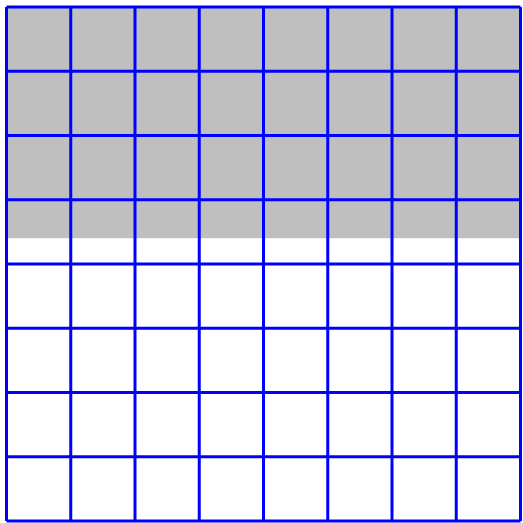}\label{fig:DomainLine}}
    \hfill
    \subfloat[][Circle]{\includegraphics[width=0.3\textwidth]{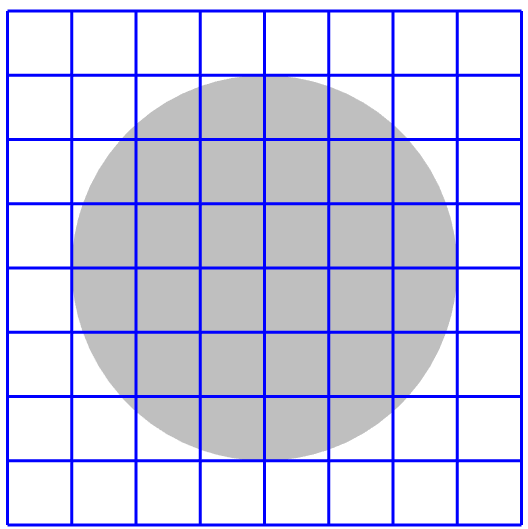}\label{fig:DomainCircle}}
    \hfill
    \subfloat[][Curvy corner cut]{\includegraphics[width=0.3\textwidth]{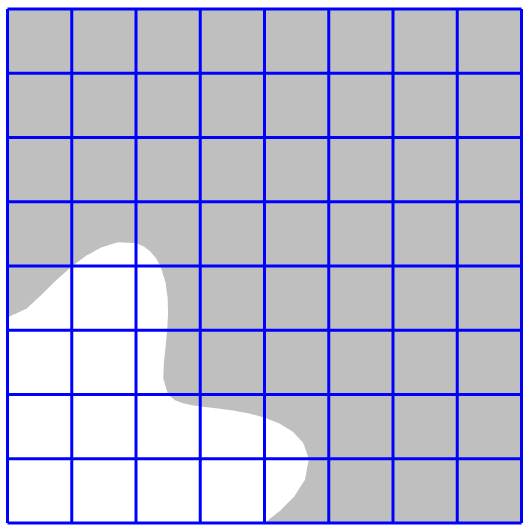}\label{fig:DomainFlowerCorner}}
    \caption{\mychange{The} different trimming cases of the numerical examples.}
    \label{fig:Domains}
\end{figure}

The numerical experiments focus on the formation of the mass matrix and $L^2$-projection for trimmed bi-variate B-splines.
\Cref{fig:Domains} illustrates the three investigated trimming cases of different complexity.

In the following, the standard element-wise Gauss procedure provides the reference solutions, and 
the fast formation and assembly strategies considered are:
\begin{itemize}
    \item \emph{\AssemblyNaive} (\cref{sec:fastformation}) for the regular support $\supportdomainReg$ of cut and interior B-splines. 
    \item \emph{\AssemblyElemLoop} (\cref{sec:HybridQuadrature}) for the regular support $\supportdomainReg$ of cut B-splines and weighted quadrature for interior B-splines. 
    \item \emph{Discontinuous weighted quadrature}  (\cref{sec:DiscontinuousWeightedQaudrature}) for the regular support $\supportdomainReg$ of cut B-splines and weighted quadrature for interior B-splines.
\end{itemize}
Cut elements are integrated by the same element-wise higher-order accurate procedure (\cref{sec:TreatmentOfCutElements}) in all cases. 
The related decomposition by Lagrange elements for the distribution of the integration point goes up to degree 6. Thus, this degree marks the upper threshold for the following numerical experiments. 

\subsection{Mass matrix formation}

The first numerical example addresses the accuracy of the formation of the mass matrix. 
Therefore, the line-trim-case (\cref{fig:DomainLine}) is considered for a 10 by 10 element basis.
Standard Gauss provides the reference matrix, and the deviation of the other matrices build by the fast concepts is measured in the Euclidean norm.
\Cref{tab:massmatrixerror} summarized the results for various degrees.
The results of the other trimming cases are omitted because they lead to the same findings and do not provide further insights.

%
\begin{table}[h]
    \centering
    \caption{The error of the computed mass matrix for the line-trim-case (\cref{fig:DomainLine}) using different fast formation and assembly concepts. 
    }
    \label{tab:massmatrixerror} 
    %
    %
    \begin{tabular}{p{1.0cm}p{2.85cm}p{2.85cm}p{2.85cm}}
        \hline\noalign{\smallskip}
        Degree & \AssemblyElemLoop & \AssemblyAddCut & \AssemblyNaive  \\
        \noalign{\smallskip}\hline\noalign{\smallskip}
        1 &	4.68883379e-17 &	4.69071638e-17 	 & 	 3.33078900e-03 \\ 	 
        2 &	2.18894391e-17 &	2.19723138e-17 	 & 	 3.38923479e-04 \\ 	 
        3 &	4.98024622e-17 &	4.98024919e-17 	 & 	 3.03838755e-04 \\ 	 
        4 &	3.80411220e-16 &	3.84739261e-16 	 & 	 1.78250728e-04 \\ 	 
        5 &	1.01851172e-15 &	1.01851512e-15 	 & 	 1.59974261e-04 \\ 	 
        6 &	2.00575179e-15 &	2.00590425e-15 	 & 	 1.28293909e-04 \\ 
        \noalign{\smallskip}\hline\noalign{\smallskip}
    \end{tabular}
\end{table}

It is apparent that the direct application of weighted quadrature to cut basis functions is not able to derive the correct mass matrix, whereas the proposed concepts are accurate up to machine precision. 
Thus, the remaining numerical examples will only employ the hybrid Gauss and discontinuous weighted quadrature approaches.

\subsection{$L^2$-projection}

Here, the accuracy and efficiency of the hybrid Gauss approach and the discontinuous weighted quadrature are investigated for all three trimming cases shown in \cref{fig:Domains}.
One ingredient for the quality of the results is the conditioning of the mass matrix, which may suffer due to the presents of trimmed basis functions. 
 The extended B-spline concept is employed for all simulations to guarantee well-conditioned mass matrices. For details on this approach, the interested reader is referred to \cite{Hoellig2003b,marussig2016a,marussig2018a,Schoellhammer2020a}.

\subsubsection{Approximation quality}

$L^2$-projection is used to fit the trimmed patch to the target function $f(x,y) = \sin(2x) \cos(3y)$, and the quality of the resulting approximation is measured in the relative $L^2$-error norm $\| \epsilon_{rel}\|_{L^2}$.
The convergence results for the line-, circle-, and curvy corner cut trimming cases are illustrated in \cref{fig:LineL2Final,fig:CircleL2Final,fig:FlowerCornerL2Final}, respectively.
In all cases, optimal convergence rates and excellent agreement to the reference solutions can be observed for both approaches proposed. 

\begin{figure}[h]
    \centering
    \tikzfig{tikz/ConvergenceTrimmedbyLineFinal.tex}{0.65}{Line convergence study}{fig:LineL2Final}{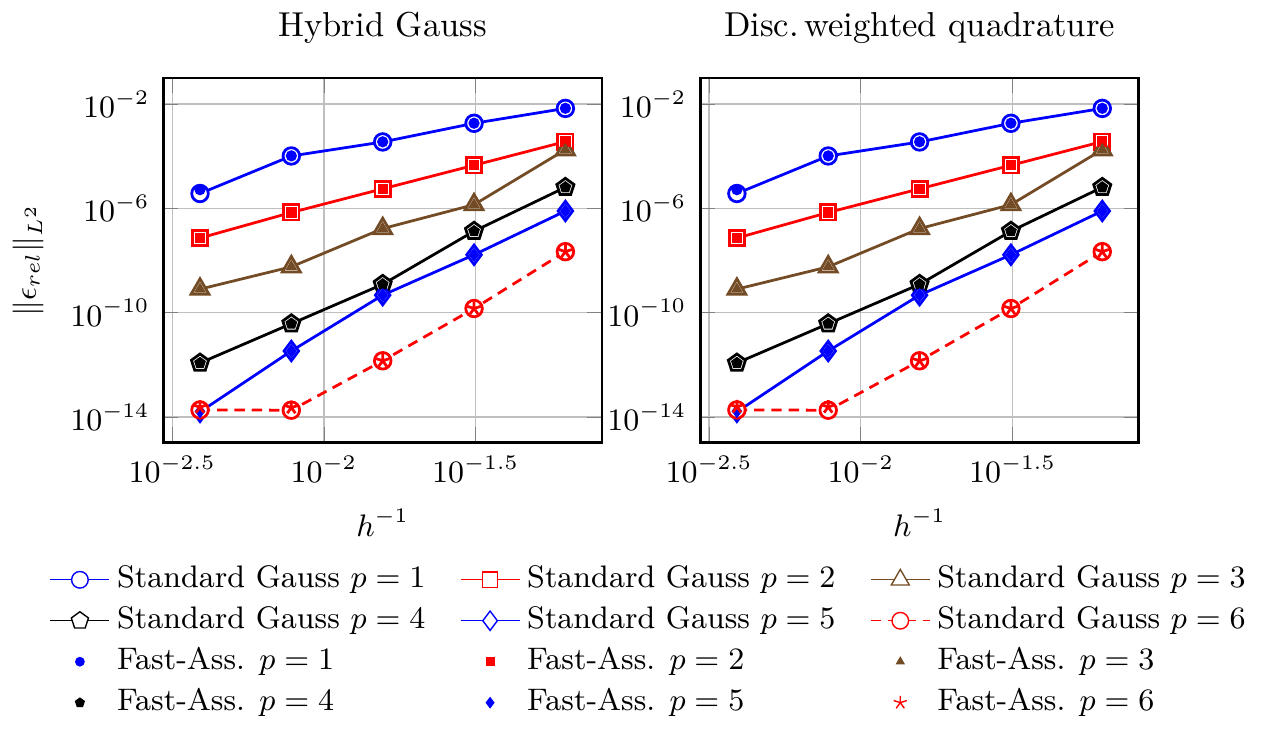}
\end{figure}

\begin{figure}
    \centering
    \tikzfig{tikz/ConvergenceTrimmedbyCircleFinal.tex}{0.65}{Circle convergence study}{fig:CircleL2Final}{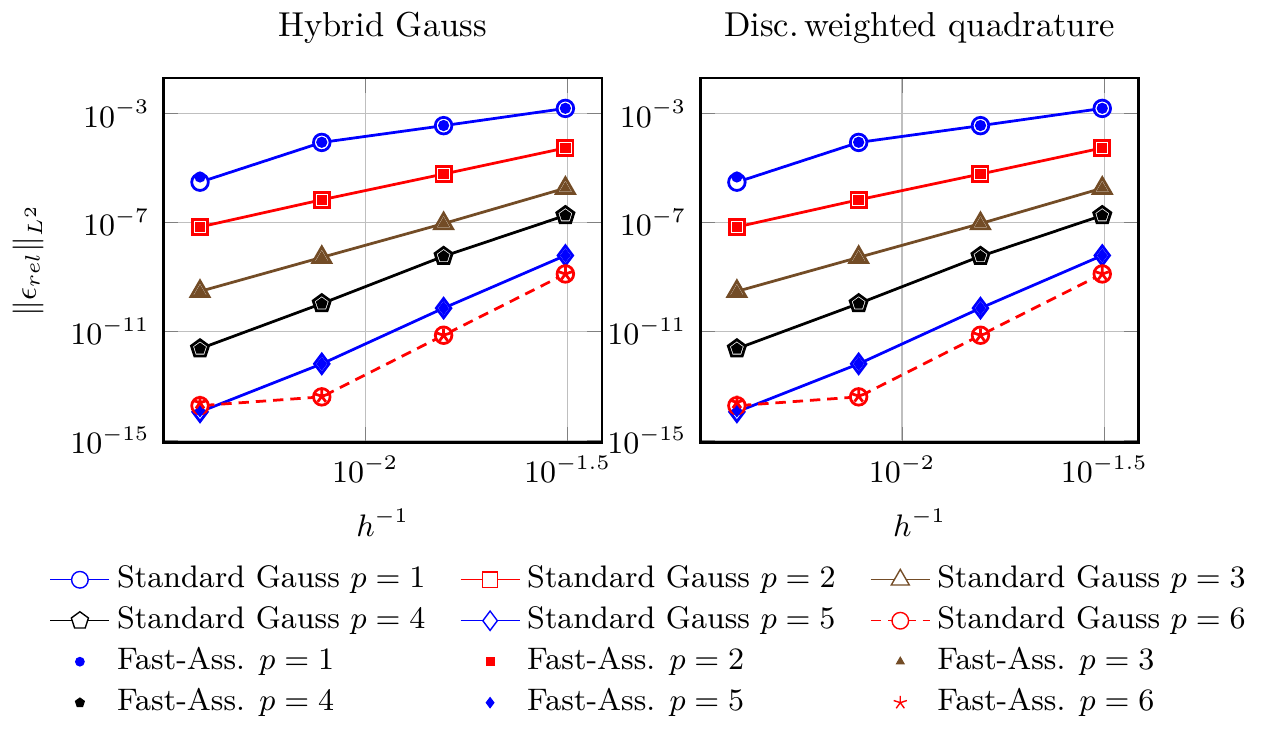}
\end{figure}
\begin{figure}
    \centering
    \tikzfig{tikz/ConvergenceTrimmedbyFlowerCornerFinal.tex}{0.65}{Curvy corner cut convergence study}{fig:FlowerCornerL2Final}{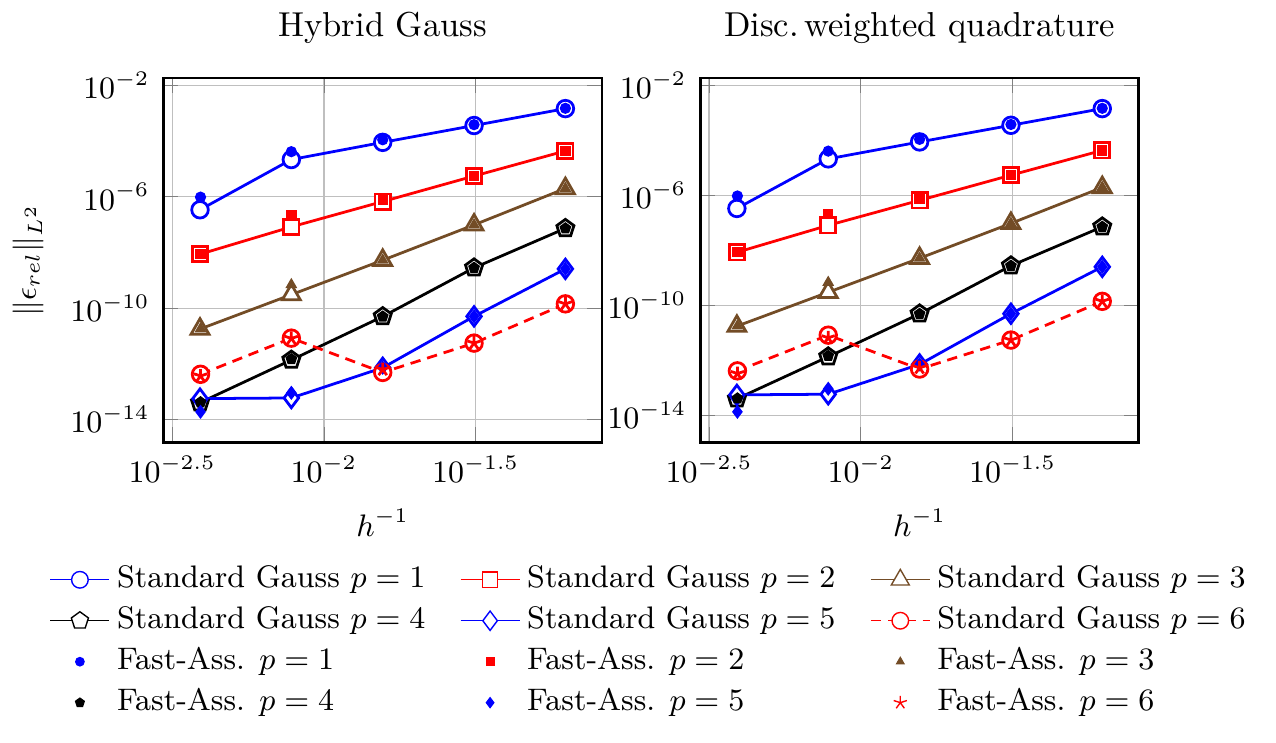}
\end{figure}

\clearpage
\subsubsection{Efficiency comparison}

The assessment of the efficiency of the different formation and assembly strategies is based on the timings of the finest discretizations used in the convergence studies.
All routines of the proposed concepts have been implemented in an in-house MATLAB\textregistered~code, which does not utilize any parallelization capabilities. 
Total timings have been obtained on an Intel\textregistered~Core\texttrademark~i7-8700 3.2~GHz processor with 16 GB RAM.
The timings have been measured with MATLAB's tic-toc command. 

\begin{figure}[b!]
    \centering
    \begin{minipage}[]{1.0\textwidth}
        \tikzsubfig{tikz/TimingsTrimmedbyLineFinal.tex}{0.65}{Line}{fig:LineFinalTime}{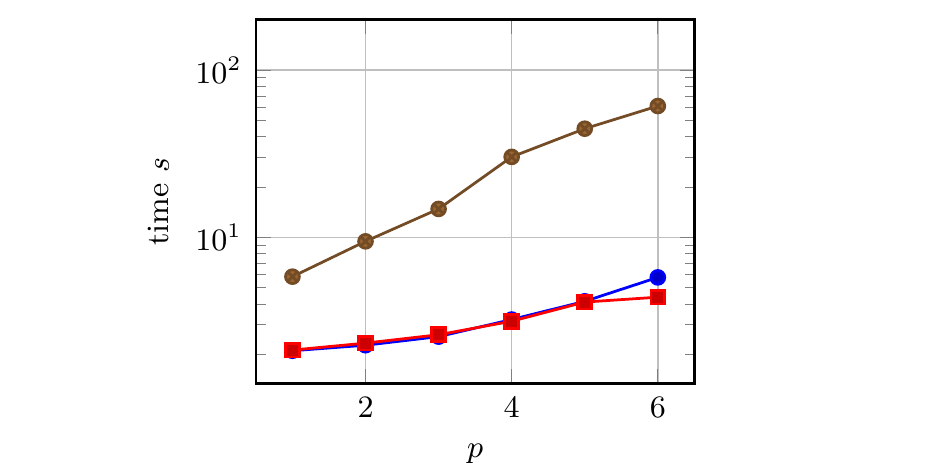}
        \vspace{0.4cm}
    \end{minipage}
    \hfill
    \begin{minipage}[]{0.45\textwidth}
        \tikzsubfig{tikz/TimingsTrimmedbyCircleFinal.tex}{0.65}{Circle}{fig:CircleFinalTime}{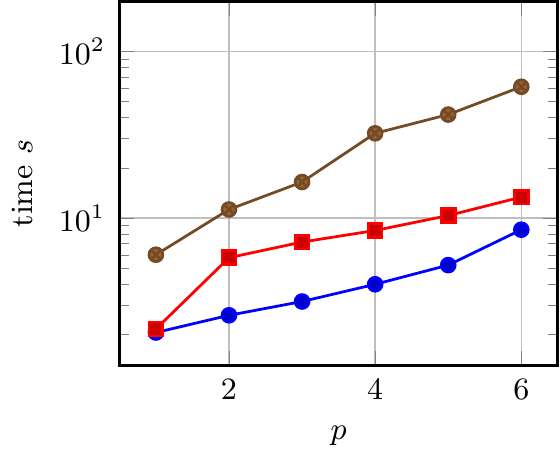}
    \end{minipage}
    \hfill
    \begin{minipage}[]{0.45\textwidth}
        \tikzsubfig{tikz/TimingsTrimmedbyFlowerCornerFinal.tex}{0.65}{Curvy corner cut}{fig:FlowerCornerFinalTime}{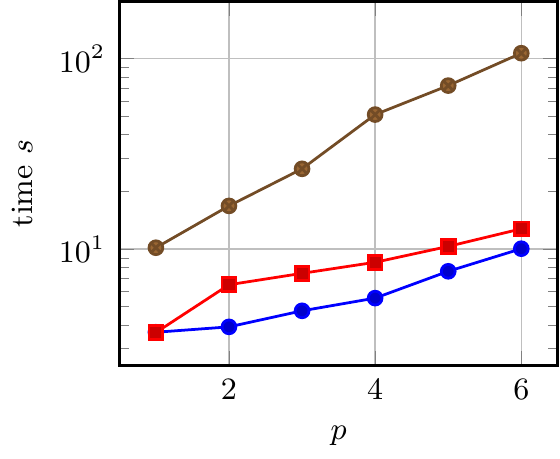}
    \end{minipage}
    \begin{minipage}[]{1.0\textwidth}
        \tikzsubfigNoCaption{tikz/TimingsTrimmedbyLineFinalLegend.tex}{0.65}{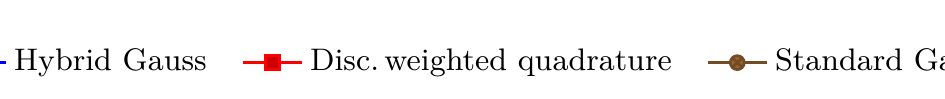}
    \end{minipage}
\caption{Total formation and assembly time including (i) the computation of quadrature weights and (ii) the integration of the entire valid support of cut and interior basis functions. Each sub-figure corresponds to a trimming case (\cref{fig:Domains}), while the different graphs refer to the assembly and formation strategy used.}
\label{fig:Timings}
\end{figure}

\Cref{fig:Timings} illustrates the total formation and assembly time obtained.
The standard Gauss formation is the most expensive concept, as expected. The performance of the other approaches, however, is a bit astonishing since the hybrid Gauss approach is more efficient in most situations.
In order to gain deeper insight, the various components of the total formation and assembly time are compared for each trimming case in \mychange{\cref{fig:TimingComponents}.}
\begin{figure}[thb]
    \centering
    \tikzfig{tikz/TimingsComponentsTrimmedAll.tex}{0.60}{The various components of the total formation and assembly time for each trimming case (\cref{fig:Domains}). The graphs refer to the set-up of the quadrature weights (Quad.~weights), as well as the integration of the regular (Interior B-splines) and trimmed ($\supportdomainReg$ of cut B-splines, Cut elements) B-splines. }{fig:TimingComponents}{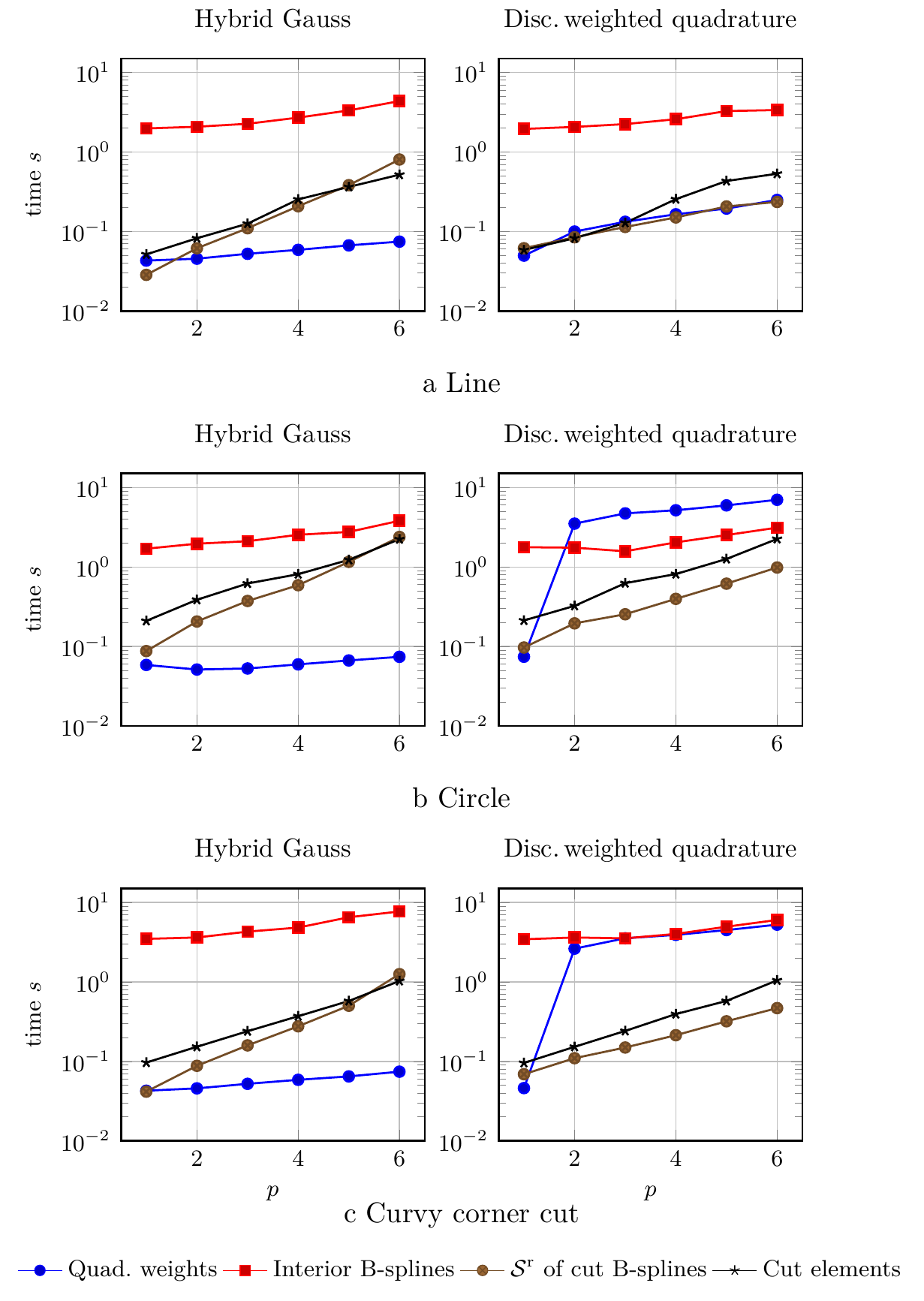}
\end{figure}
\clearpage 
These components are (i) the computation of the (normal and discontinuous) weighted quadrature rules, (ii) the formation of interior B-splines, the integration over (iii) the regular part $\supportdomainReg$ of cut B-splines and (iv) the remaining portion $\supportdomainCut$ covered by cut elements. 
The components (ii) and (iv) employ the same routines in both approaches. Hence, there are only minor deviations due to the usual time irregularities between two runs of the same code.
Looking at the timings related to the regular part $\supportdomainReg$ of cut B-splines reveals that the discontinuous weighted quadrature scales better w.r.t.~the degree.
The computation of the corresponding quadrature weights, however, is more expensive due to the additional expenses for setting up the discontinuous weighted quadrature rules for cut B-splines.
This overhead varies with the trimming situation, which determines the number of artificial discontinuous introduced, and therefore, the number of additional quadrature rule computations.

\section{Conclusions}

It has been demonstrated that the fast assembly and formation strategy presented in \cite{Calabro2017a,Hiemstra2019a} cannot be applied directly to trimmed spline spaces. 
This work examines concepts to overcome this limitation.
In particular, B-splines $\Bspline_{\boldsymbol{\indexA}}$ cut by the trimming curve $\TrimCurve$ require an adaptation of the fast procedure.
Therefore, their support is divided into a regular part $\supportdomainReg_{\boldsymbol{\indexA}}$ 
and a trimmed part $\supportdomainCut_{\boldsymbol{\indexA}}$.
The latter consists of the elements intersected by $\TrimCurve$. 
These elements do not follow the tensor product structure of the underlying spline basis and require tailored integration rules.
There are different integration techniques for this task available in the literature.
This work employs the approach presented in \cite{Fries2015a,Schoellhammer2020a},
but any scheme that allows higher-order accurate integration (favorably in an efficient manner) may be used.
This property, however, is essential. Otherwise, the integration of the cut elements outweighs the benefits of applying fast assembly and formation of the remaining basis.

The formation of the remaining contributions of a cut B-spline $\Bspline_{\boldsymbol{\indexA}}$ 
involves the integration over its $\supportdomainReg_{\boldsymbol{\indexA}}$, i.e., all non-cut elements of  $\Bspline_{\boldsymbol{\indexA}}$ enclosed by $\TrimCurve$.
This paper presents two concepts: 
The first one is a \emph{hybrid Gauss approach} which employs standard Gaussian quadrature.  
Hence, the formation of a cut B-spline utilizes standard Gauss quadrature in all elements, but the evaluations for $\supportdomainReg_{\boldsymbol{\indexA}}$ can be implemented more efficiently.
%
The second concept introduces tailored weighted quadrature rules for a cut B-spline's $\supportdomainReg_{\boldsymbol{\indexA}}$.
These \emph{discontinuous weighted quadratures} define artificial discontinuous $\discCoeff$ between elements for the construction of the quadrature weights.
The trimming curve  $\TrimCurve$ determines the locations of  $\discCoeff$.
To be precise, they mark the knots adjacent to cut elements.
Compared to normal weighted quadrature, additional quadrature points are needed next to $\discCoeff$ to fulfill the exactness condition. 
The resulting discontinuous quadrature rules allow that the coefficients on one side of $\discCoeff$ can be set to zero without affecting the contributions of the coefficients on the other side.
This property enables a straightforward application of sum factorization to a subregion of a support.
Finally, it is emphasized that interior B-splines, i.e., those not cut by $\TrimCurve$, are treated by normal weighted quadrature and sum factorization in both approaches.

The approximation quality and efficiency of the proposed approaches have been investigated by the formation of mass matrices and $L^2$-projection for bi-variate trimmed spline spaces.
It has been shown that the hybrid Gauss approach and discontinuous weighted quadrature yield the same accuracy as the reference solutions obtained by standard element-wise formation using Gauss quadrature.
Both schemes show significant speedups compared to the reference computations.
Using discontinuous weighted quadrature, computing the contribution corresponding to $\supportdomainReg_{\boldsymbol{\indexA}}$ scales better with the degree. 
However, the computational effort for setting up the additional quadrature rules dominates the total formation and assembly time, especially for complex trimming cases with multiple $\discCoeff$.
Overall, the hybrid Gauss approach turned out as the more efficient technique for the majority of the test cases considered.
From the numerical experiments, the following conclusions may be drawn: 
(i) The hybrid Gauss scheme is sufficient for moderate degrees; (ii) discontinuous weighted quadrature rules are beneficial for high degrees such as $\pu\geqslant5$, but (iii) the computation of their weights must be performed in parallel.

Future work will focus on this parallelization aspect for the computation of the discontinuous weighted quadrature rules as well as the overall fast assembly and formation scheme.
Furthermore, the efficiency of the current hybrid Gauss can be enhanced by utilizing sum factorization. 
The performance in the case of three-dimensional problems is another topic worth exploring. 
From a conceptional point of view, the extension of the concepts presented to another parametric dimension is uncomplicated.
Considering discontinuous weighted quadrature, adding another dimension merely affects the determination of proper artificial discontinuities. 
However, the implementation is much more involved due to the increased complexity of the trimming situations.
Higher-order accurate integration of trimmed tri-variate splines is already a challenge in the context of traditional finite element setting.
Solving this challenge is a prerequisite for applying fast assembly and formation.

\bibliographystyle{myplainnat}
\bibliography{FastAssembly4Trim}

\end{document}